# Noisy Teleportation

Safa Jami[1, 2]

[1]*Department of Physics, Ferdowsi University of Mashhad, 91775-1436 Mashhad, Iran.*
[2]*Department of Physics, Azad University of Mashhad, Mashhad, Iran.*

*Abstract:* Entanglement is a key resource in many quantum information applications. One of these applications is quantum teleportation.The purpose of teleportation is sending qubits across quantum channels. In general these quantum channels are noisy and therefore limit the fidelity of transmission.In this paper we consider the effect of noise on teleportation and finally find the fidelity of teleportation in presence of noise.

## I. Introduction:

Entanglement is a key resource in many quantum information applications. One of these applications is quantum teleportation [1,2]. Quantum teleportation is a technique for moving quantum state around. Alice and Bob met long ago and share an EPR pair, each taking one qubit of the pair when they separated. Now they live far apart and Alice's mission is to deliver a qubit $|\psi\rangle$ to Bob. For this purpose she interacts the qubit $|\psi\rangle$ whit her half of the EPR pair and then measure the two qubits in her possession, obtaining one of four possible classical results 00, 01, 10, 11. She sends this information to Bob. Depending on Alice's classical message Bob performs one of four operations *I*, *X*, *Z* and *ZX* on his half of the EPR pair. By doing this he can recover the original state $|\psi\rangle$. This is the standard teleportation.

## II. Noise and Imperfection:

The purpose of teleportation is sending qubits across quantum channels. In general these quantum channels are noisy and therefore limit the fidelity of transmission. The noisy channel is due to some different reasons [3]. One of them is imperfect local operations. In every real situation, the local operation applied to one or more particle will contain some imperfections. Since such operations are the building blocks for any quantum teleportation, their imperfections will limit the maximum attainable fidelity for an EPR pair. Imperfections of local operations can have various origins. To give a precise description of a noisy 1-qubit and 2-qubit operation, one needs to know the exact form of the error mechanism, which in turn depends on the specific physical implementation. In general we have only limited knowledge of these details. A generic model for a noisy channel is the so-called depolarizing channel. It transforms a qubit with initial state $\rho$ into

$$\rho \to p\rho + \frac{1-p}{2}I \tag{1}$$

where I is the identity operator and $0 \leq p \leq 1$. The action of the channel results in mixing a completely depolarized state $I/2$ to the initial density operator. The limit $p \to 0$ corresponds to a very noisy channel, while $p \to 1$ describes a channel with very little noise (ideal case). Foe example if one creates a maximally entangled state and sends each half of them to Alice and Bob, through a depolarizing channel, the result is a Werner state between Alice and Bob with fidelity F:

$$\rho_W = F|\varphi^+\rangle\langle\varphi^+| + \frac{1-F}{3}\left(|\varphi^-\rangle\langle\varphi^-| + |\psi^+\rangle\langle\psi^+| + |\psi^-\rangle\langle\psi^-|\right) \tag{2}$$

Now we consider one-qubit and two-qubit operations acting on an entangled state $\rho$ of several qubits. Let $\rho$ be a state of *n* qubits. An imperfect one-qubit operation acting on the first qubit, is described by the map

$$O_1\rho = p_1 O_1^{ideal}\rho + \frac{1-p_1}{2}tr_1\{\rho\} \otimes I_1 \tag{3}$$

whereas an imperfect two-qubit operation acting on qubit 1 and 2 is described by

$$O_{12}\rho = p_2 O_{12}^{ideal}\rho + \frac{1-p_1}{4}tr_{12}\{\rho\} \otimes I_{12} \tag{4}$$

Where $O^{ideal}$ is the ideal (perfect) operation and $I_1$, $I_{12}$ denote identity operators on the related subspace.

The other source for error and imperfection is imperfect measurement. An imperfect measurement on a single qubit is correspond to the imperfect projectors

$$\begin{aligned}P_0^\eta &= \eta|0\rangle\langle 0| + (1-\eta)|1\rangle\langle 1| \\ P_1^\eta &= \eta|1\rangle\langle 1| + (1-\eta)|0\rangle\langle 0|\end{aligned} \tag{5}$$

These expressions mean that when we try to measure 0, we obtain 0 just with probability $\mu$ and we also obtain the wrong result 1 with probability $(1-\eta) \neq 0$. That is the result is not completely reliable. An ideal measurement is described by $\eta = 1$.

### III. Fidelity of Noisy Teleportation:

After reviewing all kind of noise, now we are going to find the fidelity of teleportation in presence of noise. For doing this we must consider the effect of all imperfect operations and measurements in teleportation.

Alice wants to deliver a message $|\psi\rangle$ to Bob

$$|\psi\rangle = \alpha|0\rangle + \beta|1\rangle \tag{6}$$

As their quantum channel is noisy, they share a Werner state (2). Now Alice interacts $|\psi\rangle$ with her qubit. She does this with performing an imperfect *CNOT* on her qubit, where $|\psi\rangle$ is control and Alice's qubit is is target. Then she must perform a Bell measurement on her qubits. This measurement is imperfect. After that, she sends measurement result to Bob and Bob performs an imperfect single operation on his qubit. After tracing over Alice's qubit we have

$$\rho'_B = tr_{\psi A}(\rho') = p_1^2 p_2 \eta^2 \left(\frac{4F-1}{3}\right)|\psi\rangle\langle\psi| + p_1^2 p_2 (1-\eta)^2 \left(\frac{4F-1}{3}\right) ZX|\psi\rangle\langle\psi|XZ$$

$$+ \left[\frac{1-p_1^2 p_2}{2} + \eta(1-\eta)p_1^2 p_2 + 2p_1^2 p_2 \left(\frac{1-F}{3}\right)\left(\eta^2 + (1-\eta)^2\right)\right] I \tag{7}$$

To find the fidelity of channel we need to evaluate
$$F' = \langle\psi|\rho'_B|\psi\rangle \tag{8}$$

As $|\psi\rangle$ is an arbitrary state, we should integrate over all possible state. After this, we find

$$F' = \frac{1}{2}\left[1 + p_1^2 p_2 \left(\frac{2\eta^2 + 2\eta - 1}{3}\right)\left(\frac{4F-1}{3}\right)\right] \tag{9}$$

This expression is the fidelity of channel for quantum teleportation in presence of noise, or noisy teleportation. This means that if there is some noise, Bob won't receive the initial message exactly, but he will receive a Werner state with the above fidelity. The conditions $p_1 = p_2 = \eta = F = 1$ give us the ideal case. In this situation, as we knew, Bob will receive the initial state exactly.

**Acknowledgment**:
I would like to thank B. C. Sanders and S. Bandyopadhyay for valuable discussions about entanglement.

Email: jami@science1.um.ac.ir
[1] C. H. Bennett and S. J. Wiesner., Phys. Rev. Lett. **69**, 2881(1992).
[2] M. A. Nielsen and I. L. Chuang, *"Quantum computation and Quantum Information"*, Cambridge University Press, Cambridge, 2000.
[3] W. Dur, H. J. Briegel, J. I. Cirac and P. Zoller., Phys. Rev. A **59**(1),169 ( 1999) .